\documentclass[useAMS,usenatbib]{mn2e}
\usepackage{times}
\usepackage{graphicx}
\usepackage{subfigure}
\usepackage{psfrag}
\usepackage{amsmath}
\usepackage{amsfonts}
\usepackage{amssymb} 
\usepackage{flafter} 

\def\reff@jnl#1{{\rm#1\/}}
\def\aj{\reff@jnl{AJ}}                 
\def\araa{\reff@jnl{ARA\&A}}           
\def\apj{\reff@jnl{ApJ}}               
\def\apjl{\reff@jnl{ApJ}}              
\def\apjs{\reff@jnl{ApJS}}             
\def\ao{\reff@jnl{Appl.Optics}}        
\def\apss{\reff@jnl{Ap\&SS}}           
\def\aap{\reff@jnl{A\&A}}              
\def\aapr{\reff@jnl{A\&A~Rev.}}        
\def\aaps{\reff@jnl{A\&AS}}            
\def\azh{\reff@jnl{AZh}}               
\def\baas{\reff@jnl{BAAS}}             
\def\jrasc{\reff@jnl{JRASC}}           
\def\memras{\reff@jnl{MmRAS}}          
\def\mnras{\reff@jnl{MNRAS}}           
\def\pra{\reff@jnl{Phys.Rev.A}}        
\def\prb{\reff@jnl{Phys.Rev.B}}        
\def\prc{\reff@jnl{Phys.Rev.C}}        
\def\prd{\reff@jnl{Phys.Rev.D}}        
\def\prl{\reff@jnl{Phys.Rev.Lett}}     
\def\pasp{\reff@jnl{PASP}}             
\def\pasj{\reff@jnl{PASJ}}             
\def\qjras{\reff@jnl{QJRAS}}           
\def\skytel{\reff@jnl{S\&T}}           
\def\solphys{\reff@jnl{Solar~Phys.}}   
\def\sovast{\reff@jnl{Soviet~Ast.}}    
\def\ssr{\reff@jnl{Space~Sci.Rev.}}    
\def\zap{\reff@jnl{ZAp}}               
\def\nat{\reff@jnl{Nature}}            

\newcommand{\eg}{e.g. }

\title[F.~Feroz et al.]{Bayesian modelling of clusters of galaxies
from multi-frequency pointed Sunyaev--Zel'dovich observations}
  
\author[F.~Feroz et al.]
{Farhan Feroz\thanks{E-mail: f.feroz@mrao.cam.ac.uk},
Michael P.~Hobson,
Jonathan T.~L.~Zwart,\newauthor
Richard D.~E.~Saunders
and Keith J.~B.~Grainge
\\
Astrophysics Group, Cavendish Laboratory, J.~J.~Thomson Avenue,
Cambridge, CB3 0HE
}

\begin{document}

\date{Accepted ---. Received ---; in original form \today}
\pagerange{\pageref{firstpage}--\pageref{lastpage}}
\pubyear{2008}

\label{firstpage}
\maketitle

\begin{abstract}
\noindent We present a Bayesian approach to modelling
galaxy clusters using multi-frequency pointed observations from
telescopes that exploit the Sunyaev--Zel'dovich effect. We use the
recently developed {\sc MultiNest} technique (\citealt{feroz08};
Feroz, Hobson \& Bridges 2008) to explore the high-dimensional
parameter spaces and also to calculate the Bayesian evidence. This
permits robust parameter estimation as well as model comparison. Tests
on simulated Arcminute Microkelvin Imager observations of a cluster,
in the presence of primary CMB signal, radio point sources (detected
as well as an unresolved background) and receiver noise, show that our
algorithm is able to analyse jointly the data from six frequency
channels, sample the posterior space of the model and calculate the
Bayesian evidence very efficiently on a single processor. We also
illustrate the robustness of our detection process by applying it to a
field with radio sources and primordial CMB but no cluster, and
show that indeed no cluster is identified. The extension of our
methodology to the detection and modelling of multiple clusters in
multi-frequency SZ survey data will be described in a future work.
\end{abstract}


\section{Introduction}\label{SZ:sec:intro}

Clusters of galaxies are the most massive gravitationally bound
objects in the Universe and as such are critical tracers of the
formation of large-scale structure. The size and formation history of
massive clusters is such that the ratio of cluster gas mass to total
mass is expected to be representative of the universal ratio
$\Omega_{\rm b}/\Omega_{\rm m}$, once the relatively small amount of
baryonic matter in the cluster galaxies is taken into account (see
e.g.~\citealt{white93}). The comoving number density of clusters as a
function of mass and redshift is expected to be particularly sensitive to the
cosmological parameters $\sigma_8$ and $\Omega_{\rm m}$ (see
e.g.~\citealt{battye03}). This has been predicted both analytically (see
e.g.~\citealt{press74,sheth01}) and from large-scale numerical
simulations (see e.g.~\citealt{jenkins01,evrard02}), but cluster number densities have not yet
been measured at redshifts $z\gtrsim 1$, because of the basic problem
of the dimming of surface brightness with redshift. Moreover, optical
observations are confused by foreground galaxies, and both optical and
X-ray observations are biased towards strong mass concentrations such
as clumps.

The Sunyaev--Zel'dovich (SZ, \citealt{Sunyaev70,Sunyaev72}; see \eg
\citealt{birkinshaw99} and \citealt*{carlstrom02} for reviews) effect
produces secondary anisotropies in the cosmic microwave background
(CMB) radiation through inverse-Compton scattering from the electrons
in the hot intracluster gas (which radiates via thermal Bremsstrahlung
in the X-ray waveband), and the subsequent transfer of some of the
energy of the electrons to the low-energy photons. Pointed SZ
observations of clusters have been routine for some time (see
e.g. \citealt*{birkinshaw81,birkinshaw-plus84,uson86},
\citealt{jones93}), \citealt{grainge96}, \citeauthor{grainge02a}
\citeyear{grainge02b}a,b, \citeauthor{cotter02a}
\citeyear{cotter02b}a,b, \citealt{Grainger02}, \citealt{saunders03},
\citealt{jones05}; \citealt*{carlstrom96}, \citealt{grego00},
\citealt{patel00}, \citealt{reese00}, \citealt{joy01},
\citealt{reese02}, \citealt{bonamente06}, \citealt{laroque06}). 
By fitting simple parametric cluster models to the observed data set,
one would like to make model-dependent inferences about the cluster 
parameters, i.e. to calculate the probability distribution of these
parameters. We also wish to compare
different cluster models to enhance our astrophysical
understanding. These tasks are most conveniently carried out through
Bayesian inference.

\citet{marshall03} presented a Bayesian approach for the joint
analysis of pointed SZ and weak lensing data. They used a highly effective but
computationally intensive Markov Chain Monte Carlo sampler to explore
the high-dimensional parameter space and employed the thermodynamic
integration technique to calculate the Bayesian evidence.  We have
since extended this approach to incorporate the highly efficient
parameter-space sampling method, {\sc MultiNest}
(\citealt{feroz08,multinest}), allowing us to analyse efficiently
cluster observations in the presence of multiple radio sources. The
analysis can now be done on multi-frequency SZ data. The new
generation of SZ instruments including ACT (\citealt{ACT06}), AMI
(\citealt{zwart08}), AMiBA (\citealt{amiba06}), APEX-SZ
(\citealt{APEX06}), CARMA (\texttt{www.mmarray.org}) and SPT
(\citealt{SPT04}), all have multiple frequency channels. Our algorithm
includes a more sophisticated model for the radio sources by allowing
for their spectral indices to be non-flat.  We also take into account
the noise contribution coming from the population of faint
unsubtracted radio sources.  Furthermore, the Bayesian evidence is now
handled comprehensively and used for objective and quantitative
detection of clusters (following \citealt{hobson03}) as well as for
model selection between different cluster models. In this paper we restrict 
our focus to the analysis of pointed SZ observations, but our basic
methodology can be extended to the detection and modelling of multiple
clusters in multi-frequency SZ survey data, and will be described in a
forthcoming publication.

In Section \ref{SZ:sec:AMI} we briefly
describe the AMI telescope. In Section \ref{sec:method:bayesian} we
give an introduction to the Bayesian inference. Sections
\ref{SZ:sec:sz} and \ref{sec:datamodel} describe the SZ effect and our analysis methodology
including those features described above. In Section
\ref{SZ:sec:simulation} we apply our algorithm to simulated SZ cluster
data and we present our conclusions in Section
\ref{SZ:sec:conclusions}.

\section{The Arcminute Microkelvin Imager}\label{SZ:sec:AMI}

Although the cluster modelling method we present in this paper is
quite general in nature, a principal goal of this work has been to
develop an efficient and robust technique for analysing cluster SZ
observations made by the Arcminute Microkelvin Imager (AMI,
\citealt{rk01}, \citealt{zwart08}). This instrument consists of a pair
of interferometer arrays operating currently with six frequency
channels spanning $13.9$--$18.2$~GHz for observations on angular scales
of $30^{\prime \prime}$--$10^{\prime}$. The telescope is aimed
principally at SZ imaging of clusters of galaxies
\citep{ami-a1914}. In order to couple to the extended SZ flux from
cluster gas structures, the Small Array (SA) has a large filling
factor and thus excellent temperature sensitivity. The dominant
contaminant in SZ observations at these Rayleigh-Jeans frequencies are
the radio sources (see also Section \ref{SZ:sec:noise}). The Large
Array (LA) deliberately has higher resolution and better flux
sensitivity than the SA and, observing simultaneously with the SA both
in time and frequency, allows subtraction of such radio point
sources. SA and LA parameters are summarized in Table \ref{table:ami}.

\begin{table}
\centering
\caption{AMI technical summary.}\label{table:ami}
\begin{tabular}{{l}{c}{c}}
\hline
                           & SA                 & LA                  \\
\hline
Antenna diameter           & 3.7~m              & 12.8~m              \\
Number of antennas         & 10                 & 8                   \\
Baseline lengths (current) & 5--20~m            & 18--110~m           \\ 
Primary beam (15.7~GHz)    & $20'.1$            & $5'.5$          \\
Synthesized beam       & $\approx 3^{\prime}$ & $\approx 30^{\prime\prime}$  \\
Flux sensitivity           & 30~mJy~s$^{-1/2}$  & 3~mJy~s$^{-1/2}$    \\
Observing frequency        & \multicolumn{2}{c}{13.9--18.2{~GHz}}     \\
Bandwidth                  & \multicolumn{2}{c}{4.3{~GHz}}            \\
Number of channels         & \multicolumn{2}{c}{6}                    \\
Channel bandwidth          & \multicolumn{2}{c}{0.72{~GHz}}           \\
\hline
\end{tabular}
\end{table}

\section{Bayesian inference}\label{sec:method:bayesian}

Our cluster modelling methodology is built upon the principles of
Bayesian inference; we now give a summary of this framework. Bayesian
inference methods provide a consistent approach to the estimation of a
set of parameters $\mathbf{\Theta}$ in a model (or hypothesis) $H$ for
the data $\mathbfit{D}$. Bayes' theorem states that
\begin{equation} \Pr(\mathbf{\Theta}|\mathbfit{D}, H) =
\frac{\Pr(\mathbfit{D}|\,\mathbf{\Theta},H)
\Pr(\mathbf{\Theta}|H)}{\Pr(\mathbfit{D}|H)},
\end{equation}
where $\Pr(\mathbf{\Theta}|\mathbfit{D}, H) \equiv P(\mathbf{\Theta})$
is the posterior probability distribution of the parameters,
$\Pr(\mathbfit{D}|\mathbf{\Theta}, H) \equiv
\mathcal{L}(\mathbf{\Theta})$ is the likelihood,
$\Pr(\mathbf{\Theta}|H) \equiv \pi(\mathbf{\Theta})$ is the prior
probability distribution, and $\Pr(\mathbfit{D}|H) \equiv \mathcal{Z}$
is the Bayesian evidence.

In parameter estimation, the normalising evidence factor is usually
ignored, since it is independent of the parameters $\mathbf{\Theta}$,
and inferences are obtained by taking samples from the (unnormalised)
posterior using standard MCMC sampling methods, where at equilibrium
the chain contains a set of samples from the parameter space
distributed according to the posterior. This posterior constitutes the
complete Bayesian inference of the parameter values, and can, for
example, be marginalised over each parameter to obtain individual
parameter constraints.

In contrast to parameter estimation, for model selection the
evidence takes the central role and is simply the factor required to
normalize the posterior over $\mathbf{\Theta}$:
\begin{equation}
\mathcal{Z} =
\int{\mathcal{L}(\mathbf{\Theta})\pi(\mathbf{\Theta})}d^D\mathbf{\Theta},
\label{eq:3}
\end{equation} 
where $D$ is the dimensionality of the parameter space. As the average
of the likelihood over the prior, the evidence is larger for a model
if more of its parameter space is likely and smaller for a model with
large areas in its parameter space having low likelihood values, even
if the likelihood function is very highly peaked. Thus, the evidence
automatically implements Occam's razor: a simpler theory with compact
parameter space will have a larger evidence than a more complicated
one, unless the latter is significantly better at explaining the data.
The question of model selection between two models $H_{0}$ and $H_{1}$
can then be decided by comparing their respective posterior
probabilities, given the observed data set $\mathbfit{D}$, via the model
selection ratio $R$:
\begin{equation}
R=\frac{\Pr(H_{1}|\mathbfit{D})}{\Pr(H_{0}|\mathbfit{D})}
=\frac{\Pr(\mathbfit{D}|H_{1})\Pr(H_{1})}{\Pr(\mathbfit{D}|
H_{0})\Pr(H_{0})}
=\frac{\mathcal{Z}_1}{\mathcal{Z}_0}\frac{\Pr(H_{1})}{\Pr(H_{0})},
\label{eq:3.1}
\end{equation}
where $\Pr(H_{1})/\Pr(H_{0})$ is the \textit{a priori} probability
ratio for the two models, which can often be set to unity but
occasionally requires further consideration.

Various alternative information criteria for astrophysical model
selection are discussed by \citet{liddle07}, but the evidence remains
the preferred method. However, evaluation of the multidimensional
integral in (\ref{eq:3}) is a challenging numerical task. Standard
techniques like thermodynamic integration are extremely
computationally expensive, which makes evidence evaluation at least an
order-of-magnitude more costly than parameter estimation. Some fast
approximate methods have been used for evidence evaluation, such as
treating the posterior as a multivariate Gaussian centred at its peak
(see e.g. \citealt{HobsonBL02}), but this approximation is clearly a
poor one for multimodal posteriors (except perhaps if one performs a
separate Gaussian approximation at each mode). The Savage-Dickey
density ratio has also been proposed (see e.g. \citealt{trotta05}) as
an exact, and potentially faster, means of evaluating evidences, but
is restricted to the special case of nested hypotheses and a separable
prior on the model parameters.

The nested sampling approach, introduced by \citet{skilling04}, is a
Monte-Carlo method targeted at the efficient calculation of the
evidence, but also produces posterior inferences as a
by-product. \citet{feroz08} and \citet{multinest} built on this nested
sampling framework, and have recently introduced the {\sc MultiNest}
algorithm which is very efficient in sampling from posteriors that may
contain multiple modes and/or large (curving) degeneracies, and also
calculates the evidence. This technique has greatly reduced the
computational cost of Bayesian parameter estimation and model
selection, and is employed in this paper.

\section{Sunyaev--Zel'dovich effect from clusters}\label{SZ:sec:sz}


The primary anisotropies in the CMB are roughly one part in $10^5$ and
reflect the intrinsic non-uniformity of the matter and radiation just
before the Universe cooled sufficiently to form transparent atomic
gas. The secondary anisotropies in the CMB are due to processes
affecting the CMB after its emission. Much the most relevant for this
work is the thermal SZ effect.

The gas temperature in galaxy clusters is $10^7$--$10^8$~K. With the
radius $r$ of the cluster, the electron number density $n_{\rm e}$,
the optical depth for Thomson scattering through the centre of the
cluster is $\tau \approx 2\,r\,n_{\rm e}\,\sigma_{\rm T} \sim
10^{-2}$, where $\sigma_{\rm T}$ the Thomson scattering
cross-section. On average, energy is transferred from the electrons to
the scattered CMB photons, with the fractional energy increase
approximately equal to $k_{\rm B}T/m_{\rm e}c^2 \sim 10^{-2}$ with
$k_{\rm B}$ the Boltzmann constant, $T$ and $m_{\rm e}$ the electron
mass and temperature respectively and $c$ the speed of light. These
factors combine to give fractional CMB temperature fluctuations of order
$10^{-4}$. A full treatment (see e.g.~\citealt{birkinshaw99}) yields
the following modification to the CMB surface brightness 
in the direction of the reservoir of electrons:
\begin{equation}
\delta I_{\rm \nu} = f(\nu) y B_{\rm \nu}(T_{\rm CMB}),
\label{SZ:eq:SZ_surface_brightness}
\end{equation}
where $B_{\rm \nu}(T_{\rm CMB})$ is the blackbody spectrum at $T_{\rm
CMB} =2.726$K, while the frequency-dependent function $f(\nu)$ in the
limit of non-relativistic plasma is given as
\begin{equation}
f(\nu) = \frac{x \left[x \coth(x/2)-4\right]}{1-e^{-x}},
\label{SZ:eq:SZ_f_nu}
\end{equation}
where
\begin{equation}
x = \frac{h_{\rm p} \nu}{k_{\rm B}T_{\rm CMB}},
\end{equation}
with $h_{\rm p}$ the Planck constant. At frequencies below 217~GHz,
the SZ effect is observed as a flux decrement. The `Comptonisation'
parameter $y$ is the integral of the gas pressure along the line of
sight $l$ through the cluster:
\begin{equation}
y = \frac{\sigma_{\rm T}}{m_{\rm e}c^2} \int n_{\rm e}k_{\rm B}T\,dl.
\label{SZ:eq:SZ_comptonisation}
\end{equation}
The integral $\mathcal{Y}$ of this Comptonisation parameter over a
cluster's solid angle $d\Omega = dA/D_{\rm \theta}^2$, with $D_{\rm
  \theta}$ the angular-diameter distance to the cluster, is
proportional to the total thermal energy content of the cluster:
\begin{equation}
\mathcal{Y} =
\int y \,d\Omega =
\frac{\sigma_{\rm T}}{m_{\rm e}c^2} \int n_{\rm e}k_{\rm B}T\,dl\,d\Omega
\propto
\frac{1}{D_{\rm \theta}^2}\int n_{\rm e} T \,dV,
\label{SZ:eq:SZ_comptonisation_int}
\end{equation}
where $dV = dA\,dl$ is an element of comoving volume. The integrated SZ
surface brightness simply depends on the total cluster mass $M$, since,

\begin{equation}
\label{eqn:sz_m}
\mathcal{Y}
\propto
\frac{\left<T\right>}{D_{\rm \theta}^2}\int n_{\rm e} \,dV
\propto
\frac{M^{2/3}}{D_{\rm \theta}^2}M
\propto
\frac{M^{5/3}}{D_{\rm \theta}^2},
\end{equation}
where we have assumed that the cluster gas mass is
proportional to its total mass, i.e. $M_{\rm g} = f_{\rm g} M$.

One final property of the SZ effect is particularly significant: for
a given cluster, the SZ surface brightness is {\it independent of
redshift}. The redshift only enters via the angular-diameter distance
$D_{\rm \theta}(z)$, which at intermediate redshifts ($0.5 \la z \la
6$) is only weakly dependent on redshift for a concordance
cosmology. Therefore, an SZ survey is expected to find all
high-redshift clusters above some mass threshold with little
dependence on the redshift.

\section{Modelling interferometric SZ data}\label{sec:datamodel}

The majority of SZ observations to date have been made with
interferometers (see section \ref{SZ:sec:intro}). These instruments
have a number of advantages over single-dish telescopes
(\citealt{zwart08}, and references therein), including their relative
insensitivity to atmospheric emission (\citealt{Lay00}), lack of
required receiver stability, and the ease with which systematic errors
such as ground spill (\citealt{Watson03}) and radio-source
contamination (see e.g.~\citealt{Grainger02}) can be minimised.

Assuming a small field size, an interferometer operating at a single
frequency~$\nu$ measures samples from the complex visibility plane
$\widetilde{I}_{\rm \nu}(\mathbfit{u})$. This is given by the weighted
Fourier transform of the surface brightness $I_{\rm \nu}(\bmath{x})$,
\begin{equation}
\widetilde{I}_{\rm \nu}(\mathbfit{u}) 
= \int A_{\rm \nu}(\mathbfit{x}) I_{\rm \nu}(\mathbfit{x}) \exp(2\pi i\mathbfit{u}\cdot\mathbfit{x})~{\rm d}^2\mathbfit{x},
\label{SZ:eq:visdef}
\end{equation}
where $\mathbfit{x}$ is the position relative to the phase centre,
$A_{\rm \nu}(\mathbfit{x})$ is the (power) primary beam of the
antennas at the observing frequency $\nu$ (normalised to unity at its
peak), and $\mathbfit{u}$ is a baseline vector in units of
wavelength. Interferometers effectively measure spatial structures in
the surface brightness $\delta I_{\rm \nu}$ as the large-angular-scale
`DC' level at the centre of the $uv$-plane is never sampled.  The
positions in the $uv$-plane at which the function $\widetilde{I}_{\rm
  \nu}(\mathbfit{u})$ is sampled are determined by the physical
positions of the antennas and the direction of the field on the
sky. The samples $\mathbfit{u}_i$ lie on a series of curves, called
$uv$-tracks.

In our model, the we assume the measured interferometer visibilities
contain contributions from the cluster SZ signal, radio point sources,
primordial CMB anisotropies and instrumental noise; these
contributions are discussed below. In short, however, each
interferometer visibility is considered to consist of a signal and
generalised noise:
\begin{equation}
V_\nu(\mathbfit{u}_i)=\widetilde{I}_{\rm
  \nu}(\mathbfit{u}_i)+N_\nu(\mathbfit{u}_i),
\label{eqn:sndef}
\end{equation}
where the signal part contains the contributions from the SZ cluster
and identified radio point sources, and the generalised noise part
contains the contributions from the unresolved background of radio
point sources, primordial CMB anisotropies and instrumental noise.

\subsection{Cluster model}\label{SZ:sec:sz:models}


To determine the contribution to the visibility data from the cluster
SZ signal, one needs to calculate the Comptonisation parameter of the
cluster given in (\ref{SZ:eq:SZ_comptonisation}), for which one must
assume a cluster geometry as well as temperature and pressure
profiles for the cluster gas.

For the cluster geometry, spherically-symmetric models are a
reasonable first approximation. Ellipticity can be added easily
through a coordinate transformation (see \citealt{marshall-thesis}),
but we will not pursue that here.

The simplest gas temperature model assumes a single temperature which
is in good agreement with low-resolution X-ray emission data (see
e.g.~\citealt{Sarazin88}). One could alternatively assume a polytropic
temperature with
\begin{equation}
p_{\rm g} \propto \rho_{\rm g}^{\gamma} \Rightarrow T \propto \rho_{\rm g}^{\gamma-1},
\label{SZ:eq:SZ_T_polytropic}
\end{equation}
with $\gamma$ being the polytropic index, and $p_{\rm g}$ and
$\rho_{\rm g}$ the gas pressure and density respectively. The
polytropic model has been found to provide a good fit to simulated
clusters (see e.g.~\citealt{Komatsu01}). However, in this work we take
the cluster gas to be isothermal.

The cluster gas density is often modelled with a $\beta$-model
(\citealt{Cavaliere76,Cavaliere78}). This has the density profile
\begin{equation}
\rho_{\rm g}(r) = \frac{\rho_{\rm g}(0)}{\left[ 1+(r/r_{\rm c})^2 \right]^{\frac{3\beta}{2}}},
\label{SZ:eq:SZ_beta}
\end{equation}
where $r_{\rm c}$ is the core radius at which the profile turns over
into a region of approximately constant density, while the outer
logarithmic slope of the profile is $3\beta$.

For simplicity, we will also assume here that the cluster gas is in
hydrostatic pressure equilibrium with the total gravitational
potential $\Phi$ of the cluster, which will be dominated by the dark
matter.  Assuming spherical symmetry, the gravitational potential
$\Phi$ must thus satisfy
\begin{equation}
\frac{d\Phi}{dr} = - \frac{1}{\rho_{\rm g}}\frac{dp}{dr}.
\label{SZ:eq:SZ_hydrostatic1}
\end{equation}
Assuming the cluster gas to be an ideal gas with temperature $T$, and
its distribution to be spherically symmetric, 
(\ref{SZ:eq:SZ_hydrostatic1}) becomes
\begin{equation}
\frac{d \log \rho_{\rm g}}{d \log r} = -\frac{GM(r)\mu}{k_{\rm B}Tr},
\label{SZ:eq:SZ_hydrostatic2}
\end{equation}
where $\mu$ is the mass per particle, approximately 0.6 times the
proton mass (see \citealt{marshall03}), $G$ is the universal
gravitational constant and $M(r)$ is the total mass internal to radius
$r$. We note that an alternative approach would be to assume a dark
matter profile, such as the NFW profile \citep{navarro97}, and use
the assumption of hydrostatic equilibrium to determine the
corresponding gas density profile (\citealt{marshall03}), but we will
not pursue that here.

%
%
%

Taking together the assumptions of hydrostatic equilibrium, spherical symmetry,
a beta profile for the gas density and an isothermal temperature profile,
(\ref{SZ:eq:SZ_beta}) and (\ref{SZ:eq:SZ_hydrostatic2}) lead to
\begin{equation}
M(r) = \frac{3\beta r^3}{r_{\rm c}^2 + r^2}\frac{k_{\rm B}T}{\mu G}.
\label{SZ:eq:beta_hydostatic}
\end{equation}
This can be used to calculate the total cluster mass out to $r_{200}$, the
radius inside which the average total density is 200 times the critical
density $\rho_{\rm crit}$, as
\begin{eqnarray}
M_{200} &=& \frac{4\pi}{3}r_{200}^3 (200 \rho_{\rm crit})\nonumber\\
&=& \frac{3\beta r_{200}^3}{r_{\rm c}^2 + r_{200}^2}\frac{k_{\rm B}T}{\mu G}.
\label{SZ:eq:m200}
\end{eqnarray}
For the sake of brevity, we shall refer to $M_{200}$ simply as the
total mass of the cluster, and write it as $M$.

Although the beta model (\ref{SZ:eq:SZ_beta}) for the gas density
profile is expressed in terms of $\rho_{\rm g}(0)$, the more
interesting parameter is the total gas mass $M_{\rm g}(r)$ inside a
certain radius $r$. This can easily be accommodated by instead
treating $M_{\rm g}(r_{200})\equiv M_{\rm g}$ as a model parameter and
recovering $\rho_{\rm g}(0)$ by numerically integrating the gas
density profile to a given radius and subsequently normalizing the gas
mass within this radius. For this work, a radius of 1 $h^{-1}$Mpc was
used as the limit of integration for the gas density profile.

Thus, the parameters of our cluster model are taken to be $T$, $M_{\rm
  g}$, $\beta$ and $r_{\rm c}$, along with the position $(x_c,y_c)$
of its centroid on the sky and its redshift $z$. For a given set of
values for these parameters, a map of the predicted Comptonisation
parameter can be calculated by evaluating the line-of-sight integral
of the projected gas density profile:
\begin{equation}
y(s) = \frac{\sigma_{\rm T}}{m_{\rm e}c^2} \int_{-\infty}^{\infty} n_{\rm e}k_{\rm B}T \,dl,
\label{SZ:eq:SZ_ymap1}
\end{equation}
where $s=\theta D_{\rm \theta}$ is the projected radius, such that
$r^2 = s^2 + l^2$. Since
\begin{equation}
\rho_{\rm g} = n_{\rm e} \mu_{\rm e},
\end{equation}
where $\mu_{\rm e} = 1.14 m_{\rm p}$ (\citealt{Mason00,jones93}) is
the gas mass per electron, (\ref{SZ:eq:SZ_ymap1}) becomes
\begin{equation}
y(s) \propto \int_{-r_{\rm lim}}^{r_{\rm lim}} \rho_{\rm g}\left(r\right)T \,dl,
\label{SZ:eq:SZ_ymap2}
\end{equation}
where $r_{\rm lim}$ defines the radial limit of the integration. We
set $r_{\rm lim}$ to $20 h^{-1}$Mpc which is sufficiently large even
for low values of $\beta$.

It only remains to specify the prior $\pi(\mathbf{\Theta}_{\rm c})$ on
our cluster model parameters $\mathbf{\Theta}_{\rm c} \equiv (x_{\rm
  c},y_{\rm c},\beta,r_{\rm c},T,M_{\rm g},z)$.  Pointed SZ
observations will typically be directed towards (putative) clusters
already observed in the X-ray or optical bands.  Ideally, one would
perform a joint analysis of these data sets (see
e.g. \citealt{marshall03}) to constrain cluster parameters.  For our
present purpose of analysing SZ data alone, however, one can consider
observations in other wavebands simply to provide (joint) priors on
cluster parameters in the analysis of the subsequent SZ data.

For simplicity, we will assume throughout this paper that
the prior is separable, except in the parameters $M_{\rm
  g}$ and $z$, such that
\begin{equation}
\pi(\mathbf{\Theta}_{\rm c}) = \pi(x_{\rm c})\pi(y_{\rm
  c})\pi(\beta)\pi(r_{\rm c})\pi(T)\pi(M_{\mathrm{g}},z).
\label{eq:priordef}
\end{equation}
We use Gaussian priors on cluster position parameters, centered on the
pointing centre and with standard deviation $1$ arcmin. We adopt
uniform priors $\pi(r_{\rm c})=\mathcal{U}(0,1000)h^{-1}$kpc and
$\pi(\beta)=\mathcal{U}(0.3,1.5)$ on the cluster core radius and
$\beta$, the outer logarithmic slope of the gas density profile.  A
Gaussian prior on cluster temperature can be adopted with the mean and
standard deviation coming from X-ray or optical (from a velocity
dispersion) measurements. In the absence of any X-ray or optical
observation of the cluster, we use a uniform prior
$\pi(T)=\mathcal{U}(0,20)$~keV on the temperature.  Similarly, optical
observations could, in principle, provide a joint prior on the mass
and redshift of the cluster. In the absence of such observations, one
can instead assign the prior based upon an assumed cluster mass
function, as we now discuss.

Since it requires little further work, let us consider the very
general case in which the cluster gas fraction $f_{\rm g}$ is also
unknown, but we have some prior $\pi(f_{\rm g})$ on its value, perhaps
from previous observations (we will assume for simplicity that $f_{\rm
  g}$ is independent of redshift, although it is easy to relax this
condition also).  The gas mass of a cluster is related to its `total
mass' $M$ $(\equiv M_{200})$ by $M_{\rm g}=f_{\rm g}M$ and hence it is
straightforward to show that
\begin{equation}
\pi(M_{\rm g},z) = \int_0^1 \pi(f_{\rm g})
\pi(M=M_{\rm g}/f_{\rm
  g},z)\,\frac{df_{\rm g}}{f_{\rm g}}, 
\label{eq:mgfprior}
\end{equation}
where $\pi(M,z)$ is the joint prior on the total cluster mass and
redshift. In turn, we take the latter to be equal to some assumed
cluster mass function, $d^2n/dM\,dz$, appropriately normalised over
some ranges $M_{\rm min} < M \leq M_{\rm max}$ and $z_{\rm min} < z
\leq z_{\rm max}$. It is worth noting that, in the special case where
the cluster gas fraction is known, one simply has
$\pi(M_{\rm g},z) \propto \pi(M=M_{\rm g}/f_{\rm g},z)$.

\begin{figure}
\begin{center}
\includegraphics[width=0.8\columnwidth]{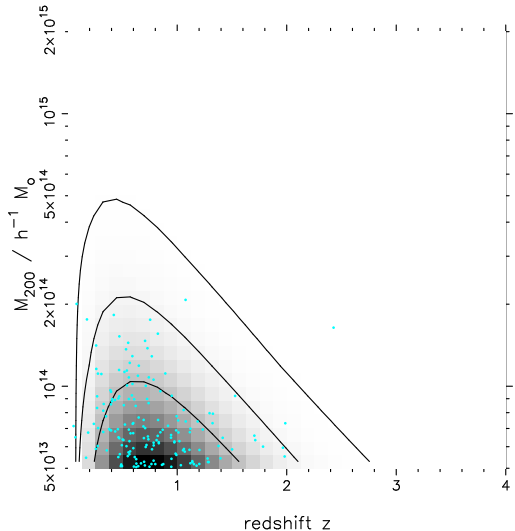}
\caption{The Press-Schechter mass function with $\sigma_8=0.8$,
together with some samples drawn from it for illustration. The
contours enclose 68\%, 95\% and 99\% of the
probability.}\label{fig:press_schechter}
\end{center}
\end{figure}

One possibility for $d^2n/dM\,dz$ is the Press--Schechter \citep{press74} mass
function. Numerical simulations have shown that the Press--Schechter
mass function overestimates the abundance of high-mass clusters and
underestimates those of low mass \citep{sheth01}, but overall it still
provides an adequate fit to $N$-body simulations \citep{jenkins01}. In
particular, a reasonable fit is obtained for the Press-Schecter mass
function with $\sigma_8=0.8$, which is plotted in
Figure~\ref{fig:press_schechter}, along with some samples drawn from
it for illustration.

Another possibility is simply to assume a separable prior in $M$
and $z$, namely $\pi(M,z) = \pi(M)\pi(z)$, where each factor 
has some simple functional form such that their product
gives a reasonable approximation to the Press--Schechter mass
function. We shall assume such a form in our analysis of simulated SZ
data in Section~\ref{SZ:sec:simulation}, where $\pi(M)$ will be taken to be
uniform in $\log M$ in the range $M_{\rm min}=10^{13} h^{-1} M_{\sun}$
to $M_{\rm max}=5 \times 10^{15} h^{-1} M_{\sun}$, and zero outside
this range.  Moreover, in Section~\ref{SZ:sec:simulation}, we will
also assume simply that the cluster redshift and gas fraction are
known, which is equivalent to imposing delta function priors on $z$
and $f_{\rm g}$.


\subsection{Resolved radio point-sources model}\label{SZ:sec:sources}

A key issue for SZ cluster observations is the effect of contaminating
radio sources. The most obvious problem is that the emission from
source(s) coincident with the cluster centre can mask the SZ
decrement. A more subtle but equally disastrous problem is that
emission from a radio source lying on a negative sidelobe of the
interferometer response (centered at the cluster) can mask an SZ
decrement. AMI's observational approach to this problem is robust:
with the RT and VSA in the past, and with AMI LA at present, we provide
enough telescope sensitivity at high angular resolution to measure the
flux densities at positions of the contaminating sources down to a
faint limiting flux density $S_{\rm lim}$. We then handle these
sources as follows.

The visibility due to each of these radio sources can be calculated
as:
\begin{equation}
\widetilde{I}_{\rm \nu}^{\rm S}(\mathbfit{u}) = \int A_{\rm \nu}(\mathbfit{x}) S_{\rm \nu}(\mathbfit{x}) \exp(2\pi
i\mathbfit{u}\cdot\mathbfit{x})~{\rm d}^2\mathbfit{x},
\label{SZ:eq:radio_source}
\end{equation}
where $S_{\rm \nu}(\mathbfit{x})$ is the source flux at a position
$\mathbfit{x}$ relative to the phase centre. Most of these foreground
radio sources appear as `point' sources in the data, in which case
the visibility can be calculated analytically:
\begin{eqnarray}
\widetilde{I}_{\rm \nu}^{\rm S}(\mathbfit{u}) &=& \int A_{\rm \nu}(\mathbfit{x}) S_{\rm \nu}
\delta(\mathbfit{x}-\mathbfit{x}_{\rm s}) \exp(2\pi i\mathbfit{u}\cdot\mathbfit{x})~{\rm d}^2 \mathbfit{x}\nonumber\\
&=& S_{\rm \nu} A_{\rm \nu}(\mathbfit{x}_{\rm s}) e^{i \phi},
\label{SZ:eq:radio_source_point}
\end{eqnarray}
where $\mathbfit{x}_{\rm s}$ is the position of the radio point source and
$\phi = 2 \pi \mathbfit{u} \cdot \mathbfit{x}_{\rm s}$ is the phase
angle. If the radio source is extended, then a Gaussian profile can be
used as a simple model for its flux distribution and
(\ref{SZ:eq:radio_source}) can again be solved analytically. For this work, however, we
assume all the radio sources to be point sources.

For SZ observations made at more than one frequency, the change of
source flux with frequency $\nu$ also needs to be taken into
account. Here we assume a power-law dependence with spectral index
$\alpha$ such that
\begin{equation}
S_\nu = S_0\left(\frac{\nu}{\nu_0}\right)^{-\alpha},
\label{SZ:eq:sz_source_spIndx}
\end{equation}
where $\nu_0$ is some reference frequency (usually that of 
the lowest observed frequency channel).

For each identified radio point source, we impose the same 
priors on its parameters $\mathbf{\Theta}_{\rm s}\equiv (x_{\rm
  s},y_{\rm s},S_0,\alpha)$.
We assume throughout that the prior is separable:
\begin{equation}
\pi(\mathbf{\Theta}_{\rm s}) = \pi(x_{\rm s})\pi(y_{\rm s})\pi(S_0)\pi(\alpha).
\label{eqn:rsprior}
\end{equation}
For the source position parameters, we use a delta function prior
centered on the measured position that is assumed known from
higher-resolution observations with the AMI LA. Although, such
observations also yield a source flux measurement that can be used to
give a very narrow prior on $S_0$, for the analysis in this paper, we
use a uniform prior $\mathcal{U}(0,20)$~mJy to allow our algorithm to
determine how accurately the source fluxes can be fitted.

\begin{figure}
\begin{center}
\includegraphics[width=10cm,origin=br,angle=0]{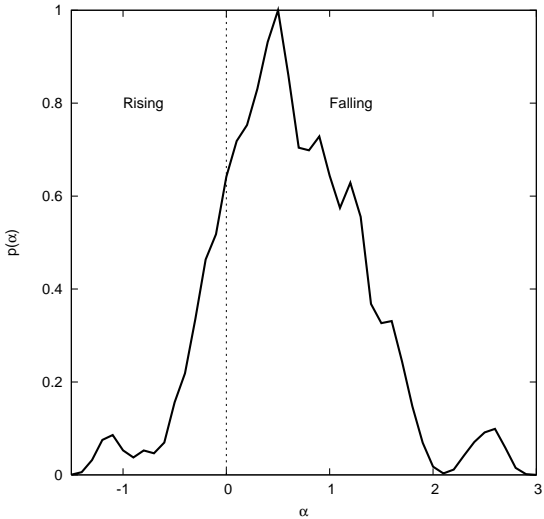}
\end{center}
\caption{Spectral index distribution from \citet{waldram07} used as
the prior on $\alpha$, normalized to unit probability at maximum. The
final normalisation depends on the adopted prior
range.}\label{SZ:fig:palpha}
\end{figure}

For the spectral index $\alpha$, a little more care is required in the
choice of prior. Although radio source spectra typically fall with an index
of 0.7, they differ widely because of synchrotron self-absorption and
ageing effects, and so can vary between $\approx 2$ (steeply falling),
through to $\approx 0$ (flat) and even $\approx -1$ (rising). Flat- and
rising-spectrum sources tend to be more variable in time, and we can
straightforwardly account for this by increasing a source's flux
uncertainty for flat or rising sources. \cite{waldram07} found 15--22
GHz spectral indices for a sample of 110 9C \citep{waldram03}
sources. The modal spectral index was found to be 0.5. We use the
distribution (Figure \ref{SZ:fig:palpha}) of these spectral indices as
the prior $\pi(\alpha)$, binned onto a grid with
$\Delta\alpha=0.1$. We expect a source's spectral index to be
well-constrained by any available long-baseline data, such as that
from the LA, and so inform the prior.


Finally, since numerous radio sources are typically identified in each
field, we combine the parameters $\mathbf{\Theta}_{\rm s}$ for each
source in to a single radio-sources parameter set $\mathbf{\Psi} =
(\mathbf{\Theta}_{\rm s1},\mathbf{\Theta}_{\rm s2},\ldots)$ and
we assume the priors on the parameters of different sources are separable.

\subsection{Generalised noise model and likelihood function}\label{SZ:sec:noise}

As mentioned previously, there are three components that contribute to
the generalised `noise' on the visibilities (\ref{eqn:sndef}):
instrumental noise, primordial CMB anisotropies and confusion noise
from the background of unsubtracted radio sources. These contributions
are discussed below and define the likelihood function for the data.

Following \citet{Hobson_Maisinger02}, it is convenient first to place
the $N_{\mathrm{vis,\nu}}$ observed complex visibilities
$V_\nu(\mathbfit{u}_i)$ in frequency channel $\nu$ into a real data
vector $\mathbfit{d}_\nu$ with components
\begin{equation}
\mathbfit{d}_{\rm \nu,i} = \left\{ 
\begin{array}{ll}
\textrm{Re}\{V_\nu(\mathbfit{u}_i)\} & \mbox{$\left(i \leq N_{\rm vis,\nu}\right)$} \\
 & \\
\textrm{Im}\{V_\nu(\mathbfit{u}_{i-N_{\rm vis,\nu}})\} & \mbox{$\left(N_{\rm vis,\nu}+1 \leq i \leq 2N_{\rm vis,\nu}\right)$}.
\end{array}
\right.
\label{SZ:eq:szdvect}
\end{equation} 
It will also be useful to define the total data vector
$\bmath{d}=\{\bmath{d}_\nu\}$, which is the concatenation of the
individual data vectors for each frequency channel $\nu$.  Similarly,
one may define the real noise vectors $\mathbfit{n}_\nu$ containing
only the noise parts $N_\nu(\mathbfit{u}_i)$ of the complex
visibilities in (\ref{eqn:sndef}), and the concatenated noise vector
$\bmath{n}$.

The instrumental noise and primordial CMB anisotropies are well
described by Gaussian processes. The background of unresolved radio
sources is, however, a Poisson process. Nonetheless, in the limit of a
large number of unresolved sources, this contribution can also be well
approximated as Gaussian (see below). Thus, in this paper, we shall
assume a Gaussian form of the likelihood function on the full
parameter set $\mathbf{\Theta}=(\mathbf{\Theta}_{\rm
  c},\mathbf{\Psi})$:
\begin{equation}
\mathcal{L}(\mathbf{\Theta}) = \frac{1}{Z_{\rm N}}
\exp(-{\textstyle\frac{1}{2}}\chi^2),
\label{eqn:genlike}
\end{equation}
where $\chi^2$ is the standard statistic quantifying the misfit
between the observed data $\bmath{d}$ and the predicted data
$\bmath{d}^{\rm p}(\mathbf{\Theta})$,
\begin{equation}
\chi^2 = \sum_{\nu,\nu'} (\bmath{d}_\nu-\bmath{d}^{\rm p}_\nu)^{\rm T}(\bmath{C}_{\nu,\nu'})^{-1}(\bmath{d}_{\nu'}
-\bmath{d}^{\rm p}_{\nu'}),
\label{eqn:chisqdef}
\end{equation}
in which $\bmath{C}_{\nu,\nu'} \equiv \langle
\bmath{n}_\nu\bmath{n}^{\rm T}_{\nu'}\rangle$ is the generalised noise
covariance matrix relating the frequency channels $\nu$ and $\nu'$.
The normalisation factor $Z_{\rm N}$ in (\ref{eqn:genlike}) is given
by
\begin{equation}
Z_{\rm N} = (2\pi)^{(2N_{\rm vis})/2}|\bmath{C}|^{1/2},
\label{eqn:likenorm}
\end{equation}
where $\bmath{C}\equiv \langle \bmath{n}\bmath{n}^{\rm T}\rangle$ and
$N_{\rm vis}$ is the total number of visibilities in all the 
frequency channels. In this work, $Z_{\rm N}$ is independent of the
model parameters $\mathbf{\Theta}$ and hence it can be safely ignored
in the analysis.

In (\ref{eqn:chisqdef}), the predicted data $\mathbfit{d}^{\rm p}_\nu$
at each frequency are a function of the model parameters
$\mathbf{\Theta}=(\mathbf{\Theta}_{\rm c},\mathbf{\Psi})$. For a given
set of parameter values, the predicted data are calculated as follows.
First, the values of the cluster parameters $\mathbf{\Theta}_{\rm c}$
are used to calculate the predicted Comptonisation map using
(\ref{SZ:eq:SZ_ymap2}), which in turn gives the cluster surface
brightness~$\delta I_{\rm \nu}$ through
(\ref{SZ:eq:SZ_surface_brightness}) and its weighted (by the primary
beam) Fourier transform (\ref{SZ:eq:visdef}), calculated on a fine
grid in the $uv$-plane. This is then sampled at the measured
$uv$-postions $\bmath{u}_i$.  Finally, the contributions from the
identified radio point sources are added to the predicted visibilities
directly using (\ref{SZ:eq:radio_source_point}).

The covariance matrices $\mathbfit{C}_{\nu,\nu'}$ in
(\ref{eqn:chisqdef}) describe the generalised noise on the observed
visibilities resulting from instrumental (receiver) noise, primordial
CMB anisotropies and the background of unresolved radio point sources.
Since, in this paper, we are assuming a background cosmology, the
covariance matrices $\mathbfit{C}_{\nu,\nu'}$ are not functions of
the model parameters $\mathbf{\Theta}$ and hence need only be
calculated once (similarly, the Cholesky decomposition of these
matrices, required for the calculation of $\chi^2$, need only be
performed once).

Assuming the three contributions to the generalised noise are
independent, the covariance matrices can be written as
\begin{equation}
\label{eqn:covar}
\mathbfit{C}_{\nu,\nu'} = \mathbfit{C}_{\nu,\nu'}^{\rm rec} +
\mathbfit{C}_{\nu,\nu'}^{\rm CMB} + \mathbfit{C}_{\nu,\nu'}^{\rm conf}.
\end{equation}
The first term on the right hand side is a diagonal matrix with
elements $\sigma^2_{\nu,i}\delta_{ij}\delta_{\nu\nu'}$, where
$\sigma_{\nu,i}$ the rms Johnson noise on the $i$th element of the
data vector $\bmath{d}_\nu$ at frequency $\nu$.  The second term
contains significant off-diagonal elements both between visibility
positions and between frequencies, and can be calculated from a given
primary CMB power spectrum $C^{\rm CMB}_\ell(\nu)$ following
\citet{Hobson_Maisinger02}; note that in intensity units the CMB
power spectrum is a function of frequency. The third term in
(\ref{eqn:covar}) is the covariance matrix of the source confusion
noise, which allows for the remaining unresolved radio sources with
flux densities less than some flux limit $S_{\rm lim}$ that have been
left after high-resolution observation and
subtraction. \cite{scheuer57} was the first to show that the flux from
such unknown sources could be considered statistically as a Poissonian
contribution (Gaussian, in the limit of large numbers of sources; see
e.g. \citealt{condon74}) to the measured signal. The response of an
interferometer to a radio source has a positive signal and both
positive and negative sidelobes.  Scheuer's work is easily modified to
the case where all sources with fluxes greater than a limiting flux
$S_{\mathrm{lim}}$ (perhaps from another telescope) have already been
subtracted. 
Assuming the unresolved 
radio sources to be randomly distributed on the sky leads to
a flat angular power spectrum for confusion noise, given by (in intensity units) 
\begin{equation}
C^{\rm conf}_{\ell}(\nu)=\int_0^{S_{\mathrm{lim}}} S^2 n_\nu(S) dS.
\end{equation}
\noindent where $S_{\mathrm{lim}}$ is the completeness limit (at, say,
5$\sigma$) of the source subtraction survey, and $n_\nu(S) \equiv
dN_\nu (> S)/dS$ is the differential source count at frequency
$\nu$ as a function of flux $S$.  In principle, one should take into
account that the unresolved radio point sources are not randomly
distributed on the sky, and may be concentrated within clusters. This
would lead to a power spectrum that was a function of $\ell$, but we
do not pursue this further here. In any case, the power spectrum
$C^{\rm conf}_{\ell}(\nu)$ can be used to construct the corresponding
confusion noise covariance matrices in the same way as for the CMB
contribution (see \citealt{Hobson_Maisinger02} for details).

In our analysis of simulated AMI data in
Section~\ref{SZ:sec:simulation}, we assume that the differential
number count $n_\nu(S)$ does not vary with frequency over the AMI band
13.9--18.2~GHz.  The deepest source counts at frequencies above
$\approx 5$~GHz are those at 15~GHz from the 9C survey
\citep{waldram03}, for which
\begin{equation}
\label{eqn:9C}
n\left(S\right)=
51 S^{-2.15}\,\mathrm{Jy}^{-1}\,\mathrm{sr}^{-1},
\end{equation}
based on 465 sources above a 5-$\sigma$ completeness of 25~mJy. We use
this count in our analysis.

\subsection{Estimation of model parameters}\label{sec:clustpars}

The posterior $P(\mathbf{\Theta}) \propto
L(\mathbf{\Theta})\pi(\mathbf{\Theta}) $ of the model parameters
$\mathbf{\Theta}=(\mathbf{\Theta}_{\rm c},\mathbf{\Psi})$ can be
efficiently and robustly explored using the posterior weighted samples
produced by the {\sc MultiNest} algorithm
\citep{feroz08,multinest}. From these samples, one can, for example,
construct one-dimensional marginalised posterior distributions for
each parameter, from which best-fit values and uncertainties are
trivially obtained. In terms of cluster modelling, one is interested
only in the cluster parameters $\mathbf{\Theta}_{\rm c}=(x_{\rm
  c},y_{\rm c},\beta,r_{\rm c},T,M_{\mathrm{g}},z)$, whereas the parameters
$\mathbf{\Psi}$ associated with the resolved radio point sources are
considered as nuisance parameters and are marginalised over. It may,
however, also be of interest instead to marginalised over the cluster
parameters and produce one-dimensional marginals for the flux $S_0$ of
each resolved radio source, as well as its spectral index $\alpha$.

\subsection{Quantification of cluster detection}\label{SZ:sec:quantification}
\label{sec:quant}

Owing primarily to the presence of primary CMB anisotropies, it is
extremely important to quantify SZ cluster detection. We now discuss
how one may calculate the probability that the observed field does
indeed contain a real cluster above some particular mass limit of
interest.

This quantification is most naturally performed via a Bayesian model
selection by evaluating the evidence associated with the posterior for
competing models for the data (see e.g. \citealt{hobson03}). It is
convenient to consider the following models (or hypotheses):
\smallskip\newline
$H_0 = $ `a cluster with $M_{\rm g,min} < M_{\rm g}  \leq M_{\rm g,lim}$ is centred in $S$',
\smallskip\newline
$H_1 =$ `a cluster with $M_{\rm g,lim}< M_{\rm g} < M_{\rm g,max}$ is centred in $S$',
\smallskip\newline where $S$ is the total prior region in the spatial
subspace $\bmath{x}_c=(x_{\rm c},y_{\rm c})$.  Here $M_{\rm g,min}$ is
the lower limit of our assumed prior range on the cluster gas mass;
hence clusters below this minimum mass are supposed not to
exist. Similarly, $M_{\rm g,max}$ is the upper limit of our assumed
prior range. Finally, $M_{\rm g,lim}$ is the limiting gas mass of
interest that we discuss in more detail below. 

We must calculate the model selection ratio $R$ given in
(\ref{eq:3.1}) between the hypotheses $H_0$ and $H_1$.  
For each hypothesis $H_i$ $(i=0,1)$, the evidence is given by
\begin{equation}
\mathcal{Z}_i = \int
\mathcal{L}(\mathbf{\Theta})\pi_i(\mathbf{\Theta})\,d\mathbf{\Theta},
\label{eq:evid}
\end{equation}
where 
\begin{equation}
\pi_i(\mathbf{\Theta})=\pi_i(\bmath{x}_{\rm
c})\pi_i(\beta)\pi_i(r_{\rm
c})\pi_i(T)\pi_i(M_{\rm g},z)\pi_i(\mathbf{\Psi}),
\end{equation}
for $i=0,1$, are priors that define the hypotheses. In particular, the
priors on all the cluster parameters and source parameters, apart from
$M_{\rm g}$ and $z$, may be taken to be the same as those discussed
above for both hypotheses. Differences between the priors for the two
hypotheses do occur in $\pi_i(M_{\rm g},z)$, but in a
straightforward manner.  For hypothesis $H_0$, we use $\pi(M_{\rm
  g},z)$ given in (\ref{eq:mgfprior}), but now appropriately
normalised over the range $M_{\rm
  g,min} < M_{\rm g} \leq M_{\rm g,lim}$, and the prior is zero
outside this range.  Similarly, for hypothesis $H_1$, we use
(\ref{eq:mgfprior}) appropriately normalised over the range $M_{\rm
  g,lim} < M_{\rm g} < M_{\rm g,max}$, and the prior is zero outside
this range. The evidences (\ref{eq:evid}) for $i=0,1$ are easily
obtained using the {\sc MultiNest} algorithm.

So far we have not addressed the prior ratio $\Pr(H_1)/\Pr(H_0)$ in
(\ref{eq:3.1}).  This is easily obtained from the
prior distribution $\pi(M_{\rm  g},z)$ in (\ref{eq:mgfprior}), and is
given by
\begin{equation}
\frac{\Pr(H_1)}{\Pr(H_0)} = \frac{
\int_{z_{\rm min}}^{z_{\rm max}} \int_{M_{\rm
g,lim}}^{M_{\rm g,max}} \pi(M_{\rm g},z) \,dM\,dz}
{\int_{z_{\rm min}}^{z_{\rm max}} \int_{M_{\rm
g,min}}^{M_{\rm g,lim}} \pi(M_{\rm g},z) \,dM\,dz}.
\end{equation}
It is worth noting that, in the case where the cluster gas fraction
$f_{\rm g}$ is assumed known, the above prior ratio is simply
\begin{equation}
\frac{\Pr(H_1)}{\Pr(H_0)} = \frac{
\int_{z_{\rm min}}^{z_{\rm max}} \int_{M_{\rm
g,lim}/f_{\rm g}}^{M_{\rm g,max}/f_{\rm g}} \frac{d^{2}n}{dMdz} \,dM\,dz}
{\int_{z_{\rm min}}^{z_{\rm max}} \int_{M_{\rm
g,min}/f_{\rm g}}^{M_{\rm g,lim}/f_{\rm g}} \frac{d^{2}n}{dMdz}\,dM\,dz},
\label{eq:fgfix}
\end{equation}
where $d^{2}n/dMdz$ is the assumed cluster mass function, i.e.  the
distribution of the projected number density of clusters in a given
mass and redshift bin per unit area. Moreover, if it is also assumed
that the cluster redshift is known to be $z=z_{\rm c}$, 
then (\ref{eq:fgfix}) reduces to
\begin{equation}
\frac{\Pr(H_1)}{\Pr(H_0)} = \frac{
\int_{M_{\rm
g,lim}/f_{\rm g}}^{M_{\rm g,max}/f_{\rm g}}
\frac{dn}{dM}\big|_{z=z_{\rm c}} \,dM}
{\int_{M_{\rm
g,min}/f_{\rm g}}^{M_{\rm g,lim}/f_{\rm g}}
  \frac{dn}{dM}\big|_{z=z_{\rm c}}
\,dM}.
\label{eq:fgzfix}
\end{equation}
%


We are thus able to calculate the model selection ratio $R$ in
(\ref{eq:3.1}), which gives us the relative probability that the field
contains a `true' cluster, with gas mass above the limit $M_{\rm g,lim}$, 
as opposed to `false' cluster, with gas mass below this limit.
This, in turn, allows us to calculate the probability that the field
contains a `true' cluster, which is given by
\begin{equation} 
p = \frac{R}{1+R}.
\label{eq:prob_TP}
\end{equation}
%

\section{Application to simulated SZ observation}\label{SZ:sec:simulation}

\begin{figure*}
\centering
\includegraphics[height=6cm,origin=br,angle=0]{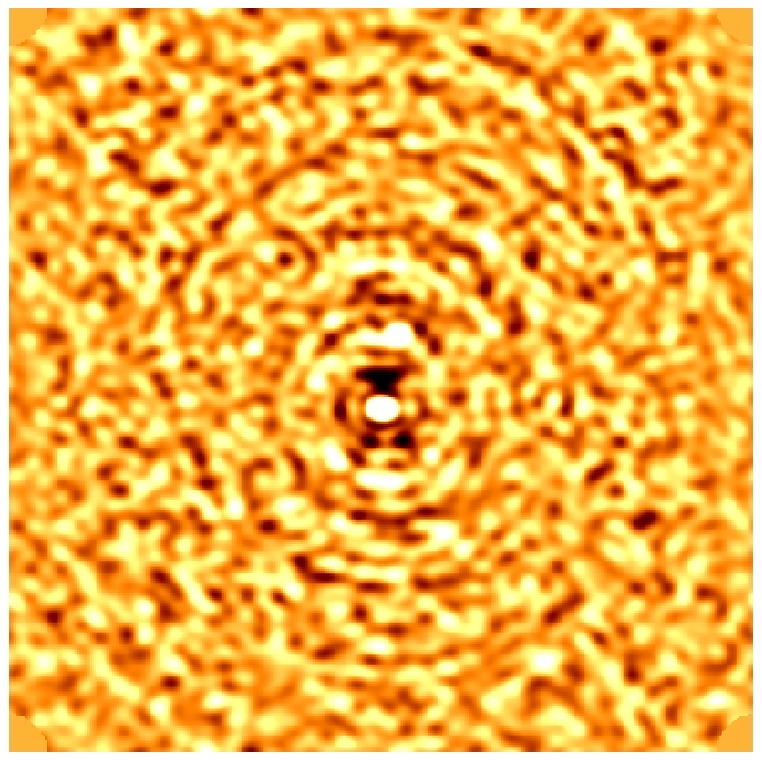}\qquad
\includegraphics[height=6cm,origin=br,angle=0]{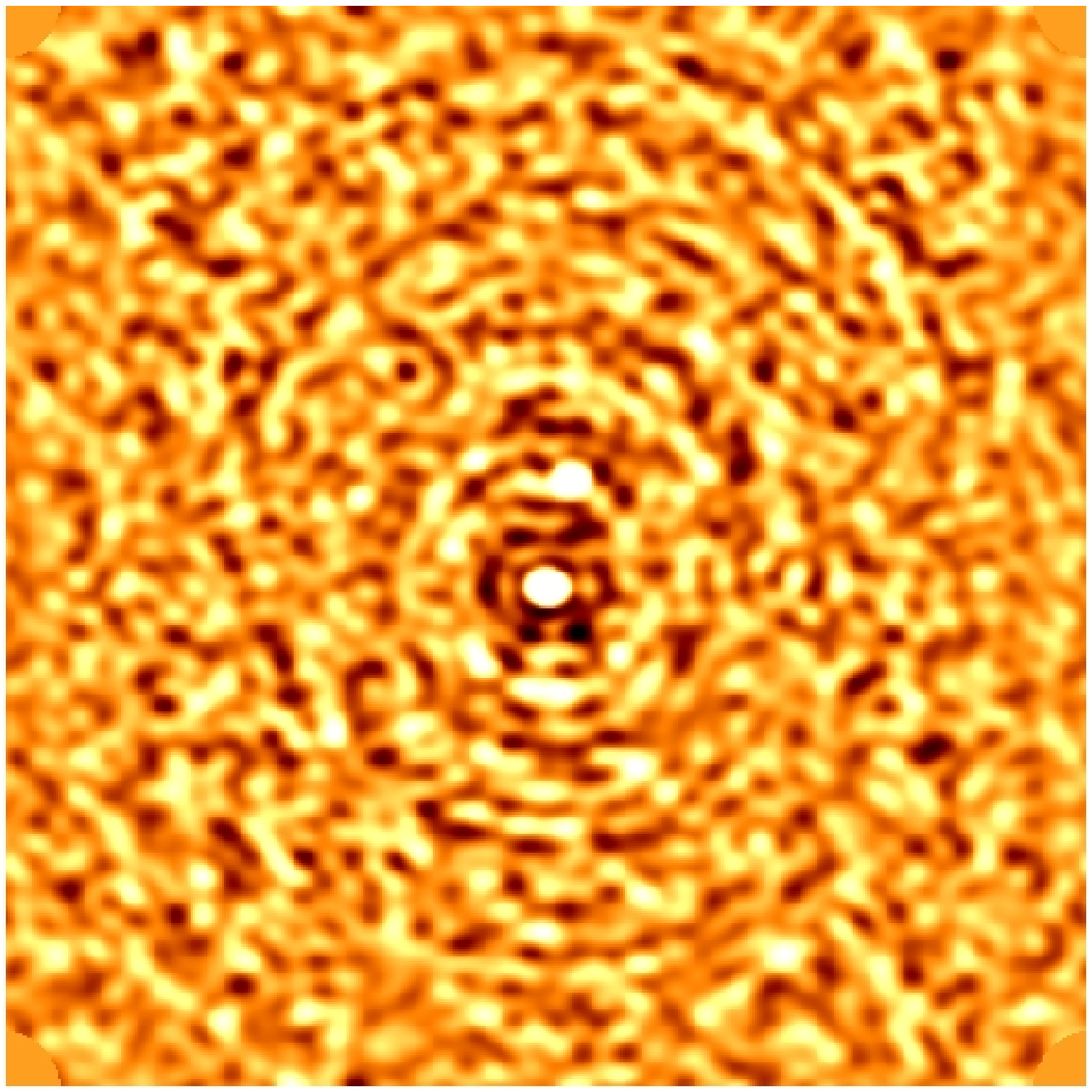}\\
\includegraphics[height=1cm,width=12.5cm,origin=br,angle=0]{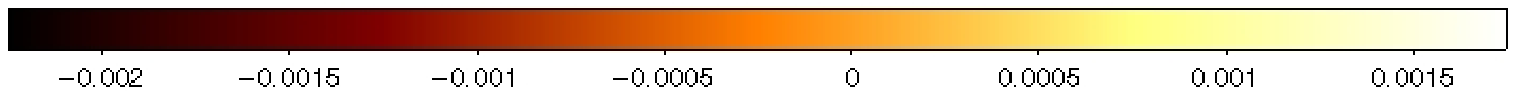}\\
\caption{Maps made from channel 4 (14.992~GHz) of the simulations
considered in this work with Simulation `A' in the left panel and
Simulation `B' in the right panel. The un\textsc{CLEAN}ed maps are
$512\times512$ pixels at $15^{\prime \prime}$ resolution. Both
simulations have the same realisation of primary CMB and instrumental
noise. The three radio sources are also the same between the two simulations.
Simulation `A' has a spherically-symmetric, isothermal
$\beta$-model cluster at the centre of the map. Simulation `B' has no
cluster.}\label{SZ:fig:sim}
\end{figure*}

In this section we describe the results of our SZ cluster modelling
algorithm on simulated SZ cluster data-sets from the Arcminute
Microkelvin Imager (AMI).

\subsection{Simulated AMI data-sets}

In simulating mock skies
and observing them with a model AMI SA, we have used the methods
outlined in \cite{Hobson_Maisinger02} and \cite{grainge02b}. 
We consider two simulations. Simulation `A' has a cluster
at $z=0.3$ modelled as a spherically-symmetric isothermal
$\beta$-profile with $r_c=60^{\prime \prime}$, $\beta=0.65$, $n_{\rm
e}=10^{-2}\,\mathrm{cm}^{-3}$ and $T=8$~keV. The gas profile is linearly
tapered to zero between $20 r_c$ and $20.01 r_c$.  The Comptonisation
$y$-parameter for this model is evaluated on a cube whose face has
$512\times 512$ pixels at $30^{\prime \prime}$ resolution before being
integrated along the line of sight. Radio point sources are added to
these maps using the fluxes, positions and spectral indices given in
Table \ref{sz:tab:sim}.

\begin{table}
\centering
\begin{tabular}{rllll}
\hline
& $\Delta x$/arcsec & $\Delta y$/arcsec & $S_{15}$/mJy & $\alpha$\\
\hline
1          & 8   & 10  & 5  & 0              \\ 
2          & 0   & --5 & 15 & +1 (falling)   \\ 
3          & --3 & 8   & 8  & --0.3 (rising) \\ 
\hline
\end{tabular}
\caption{Contaminating radio sources for the simulations considered in
this work. Source positions are given in arcminute offsets from the
pointing centre. The flux and spectral index are at
15.0~GHz.}\label{sz:tab:sim}
\end{table}

The $uv$-positions of visibility points are simulated by calculating the 
baselines (assuming SA antenna positions in \citealt{zwart08}) for a
target at right ascension $\alpha=4$~hours and declination
$\delta=+40^o$ observed over hour angle $\pm4$~hours with one-second
sampling.

For each simulation, a realisation of the primary CMB is calculated
using a power spectrum of primary anisotropies was generated for
$\ell<8000$ using CAMB \citep{camb}, with a $\Lambda$CDM cosmology
($\Omega_m=0.3$, $\Omega_{\Lambda}=0.7$, $\sigma_8=0.8$ and $h=0.7$)
assumed.
Primary CMB modes on $\ell$ scales outside the range measurable
($\ell\approx 500$--$8000$, considering the most extreme frequency
channels) by the SA are set to zero. The CMB realisation is co-added
to the cluster and radio source map in brightness temperature. 
To each model sky we also add a population of faint, confusing radio
point sources, uniformly distributed on the sky but drawn from a
Poisson distribution in flux, with the 9C source count (see section
\ref{SZ:sec:noise}) between $10\,\mu$Jy and $S_{\mathrm{lim}}=200\,\mu$Jy. The map
is scaled by the primary beam appropriate to the measured value in
that frequency channel and transformed into the Fourier plane
(equivalent to Fourier transforming and convolving with the aperture
illumination function). The resulting function is sampled at the
required visibility points and thermal receiver noise, appropriate to
the measured sensitivity of the SA, is added at this stage.

The whole process above is repeated for each of the six frequency
channels. Simulation `B' is identical to Simulation `A', but has no
cluster. Maps made from the simulated visibilities for channel 4
(14.992~GHz), for both models, are shown in Figure \ref{SZ:fig:sim}.

\subsection{Analysis and results}\label{SZ:sec:simulation:results}

%
\begin{table}
\begin{center}
\begin{tabular}{cc}
\hline
Parameters & Priors \\
\hline
$x_c,y_c$ & $(0 \pm 60)^{\prime\prime}$ \\
$M_{\rm g}$ & $12 < \log_{10} M_{\rm g}/h^{-2}M_{\sun} < 14.5$ \\
$T$ & $0 < T/\mathrm{keV} < 20$ \\
$r_c$ & $0 < r_c/h^{-1}$ kpc $< 1000$ \\
$\beta$ & $0.3 < \beta < 1.5$ \\
$S_0$ & $0 <$ $S$/mJy $< 20$ \\
\hline
\end{tabular}
\caption{Priors for the cluster and source parameters. Inequalities
denote uniform prior probability between the given limits, whilst $(a
\pm b)$ denotes a Gaussian prior with mean a and variance
$b^2$.}\label{tab:cluster_priors}
\end{center}
\end{table}

\begin{figure*}
\begin{center}
\includegraphics[width=1.8\columnwidth]{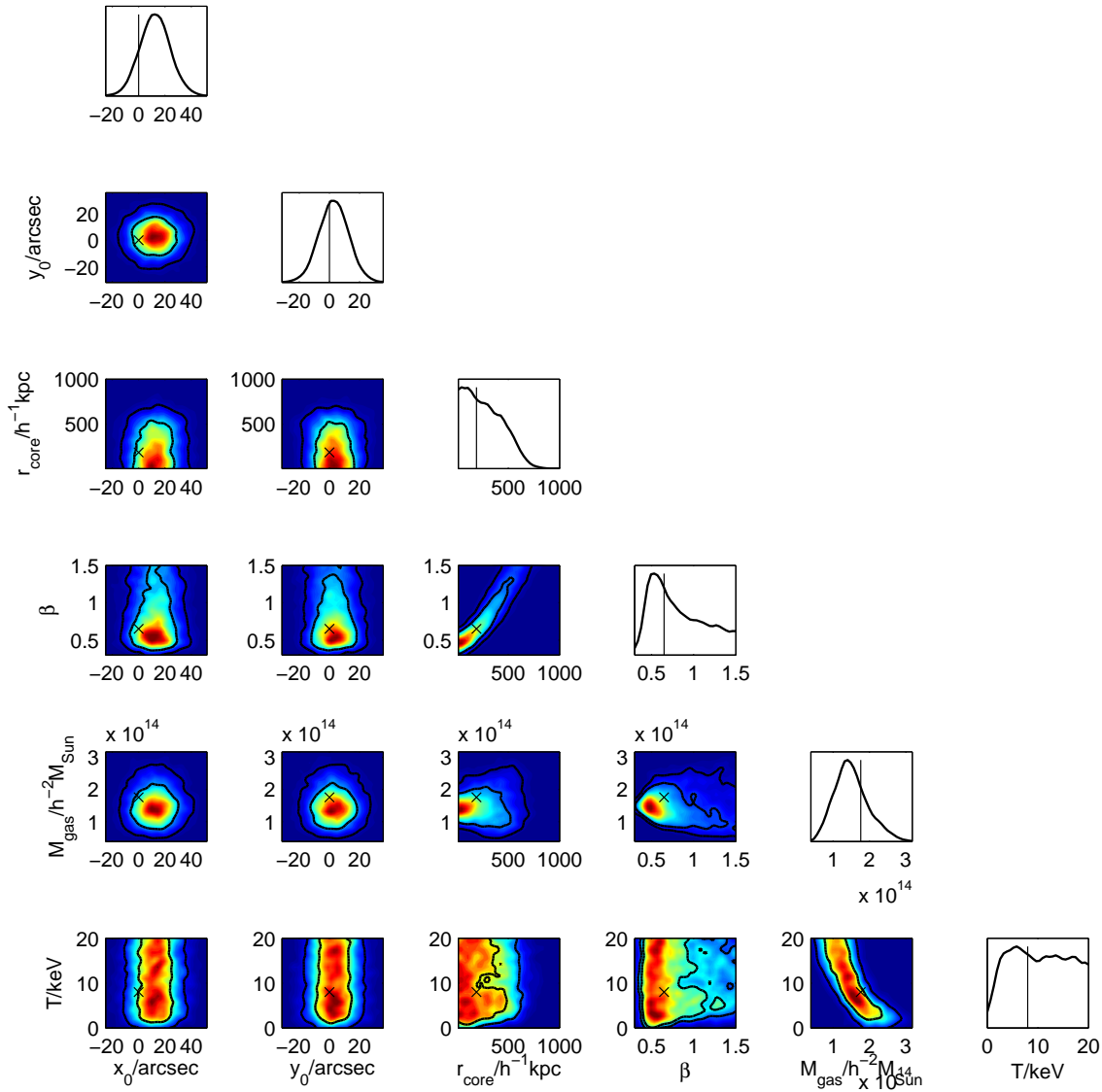}
\caption{2-D marginalized posterior probability distributions for the
parameters of the cluster Simulation `A' discussed in Section
\ref{SZ:sec:simulation}. The true parameter values used in the
simulation are shown by crosses and vertical lines in 2-D and 1-D
marginalisations respectively.}\label{SZ:fig:BL_tri}
\end{center}
\end{figure*}

We analysed the cluster simulations discussed above assuming a cluster
model with spherical geometry, a beta profile for the gas and
isothermal temperature. The priors used are listed for convenience in Table
\ref{tab:cluster_priors}. Positions of the radio point sources were
fixed to their true values. The primordial CMB and confusion noise
were included through the covariance matrix as discussed in
Section~\ref{SZ:sec:noise}. For the confusion noise we used the 9C
source count with limiting flux $S_{\rm lim}=200\mu$Jy (correctly) and
the count cut off below $10\,\mu$Jy. For the primordial CMB
anisotropies, we (correctly) assumed a $\Lambda$CDM cosmology with $\Omega_m=0.3$,
$\Omega_{\Lambda}=0.7$, $\sigma_8=0.8$ and $h=0.7$. We also assumed
that the redshift of the cluster $z=0.3$ and the gas fraction $f_{\rm
  g}=0.1$ were known.  We analysed all six AMI frequencies channels
jointly. In quantifying our cluster detection, we adopted a gas mass
limit of $M_{\rm g,lim} = 10^{13} h^{-1} M_{\sun}$ and assumed
a Press--Schechter mass function.

\begin{figure*}
\psfrag{x}[c][c]{$x/$arcsec}
\psfrag{y}[c][c]{$y$/arcsec}
\psfrag{rc}[c][c]{$r_c/h^{-1}$kpc}
\psfrag{beta}[c][c]{$\beta$}
\psfrag{T}[c][c]{$T/$keV}
\psfrag{Mgas}[c][c]{$M_{\rm g}/h^{-2}M_{\sun}$}
\psfrag{S1}[c][c]{$S_1$/mJy}
\psfrag{S2}[c][c]{$S_2$/mJy}
\psfrag{S3}[c][c]{$S_3$/mJy}
\psfrag{a1}[c][c]{$\alpha_1$}
\psfrag{a2}[c][c]{$\alpha_2$}
\psfrag{a3}[c][c]{$\alpha_3$}
\begin{center}
\includegraphics[width=1\columnwidth,angle=-90]{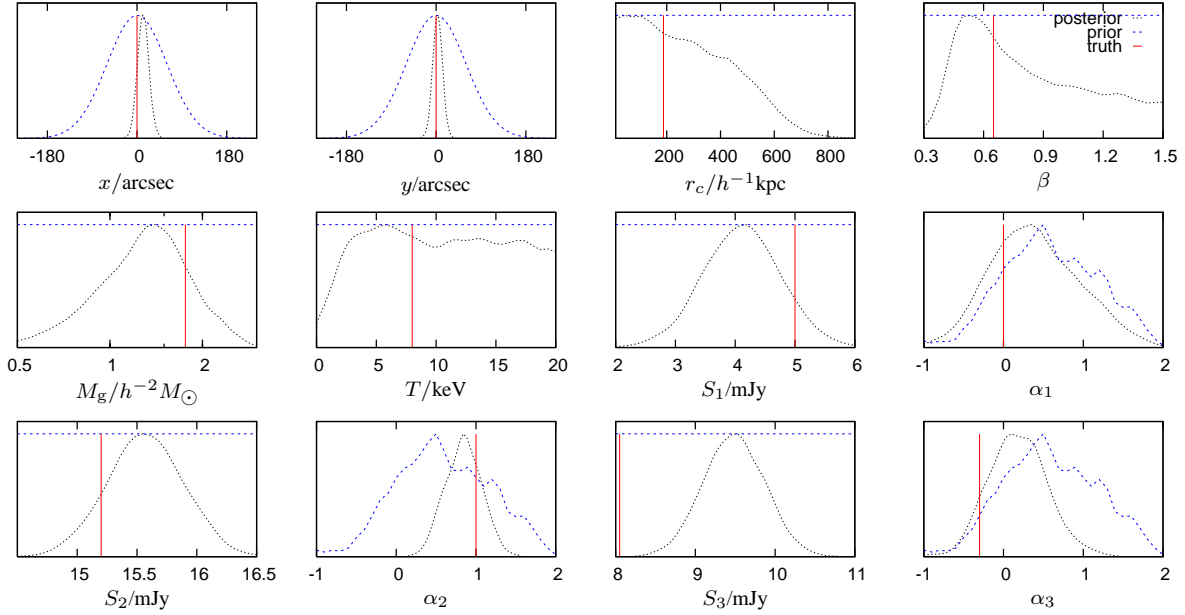}
\caption{Priors and 1-D marginalized posterior probability
distributions for the parameters of the simulated cluster and radio
point sources discussed in Section \ref{SZ:sec:simulation}. The true
parameter values used in the simulation are shown by vertical
lines.}\label{SZ:fig:BL_1d}
\end{center}
\end{figure*}

For the Simulation `B' (with radio point sources and no cluster) {\sc
  MultiNest} did identify a dominant peak in the posterior
distribution of cluster parameters but the probability odds ratio $R$,
as discussed in Section \ref{SZ:sec:quantification}, was evaluated to
be $0.32 \pm 0.03$, showing that it is more than twice as likely that
the field did not contain a `true' cluster with a gas mass above the
mass limit of interest.  Since there is no cluster in the field, the
highest likelihood point comes from a large negative
primordial feature, but since the statistics of the primordial CMB
have been incorporated in the likelihood evaluation through the
covariance matrix, the Bayesian model selection takes this into
account and consequently the odds ratio is in favour of the detected
feature being `false'. To verify this assertion, we analysed
Simulation `B' without including the CMB component in the covariance
matrix, in which case the probability odds ratio for cluster detection
$R$ was evaluated to be $\approx 150$ which clearly shows that
including the CMB is extremely important for properly modelling galaxy
clusters through the SZ effect.

\begin{figure}
\begin{center}
\includegraphics[width=1\columnwidth]{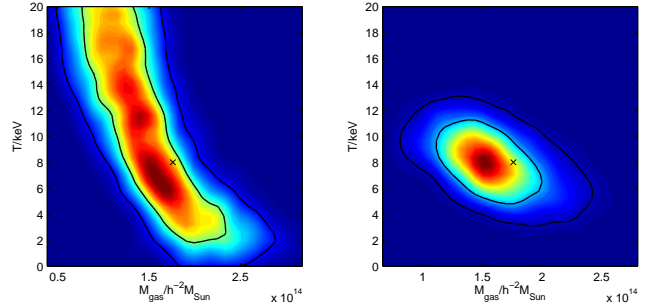}
\caption{2-D marginalized posterior probability distributions for $T$
and $M_{\rm g}$ for the analysis of the cluster Simulation `A'
discussed in Section \ref{SZ:sec:simulation}. The true parameter
values used in the simulation are shown by crosses. Uniform
$\mathcal{U}(0,20)$ keV and Gaussian $(8 \pm 2)$ keV priors on $T$
were used for the figures on the left and right hand-panels
respectively.}\label{SZ:fig:BL_UG_2d}
\end{center}
\end{figure}

\begin{figure}
\psfrag{U}[r][t]{$T=\mathcal{U}(0,20)$ keV}
\psfrag{G}[r][b]{$T=(8 \pm 2)$ keV}
\psfrag{Mgas}[c][c]{$M_{\rm g}/10^{14}h^{-2}M_{\sun}$}
\begin{center}
\includegraphics[width=0.6\columnwidth,angle=-90]{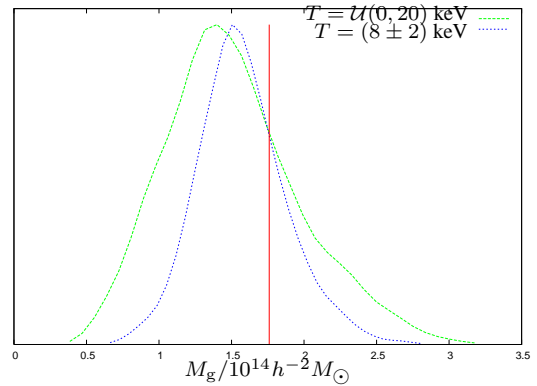}
\caption{1-D marginalized posterior probability distributions for
$M_{\rm g}$ for the analysis of the cluster Simulation `A' discussed
in Section \ref{SZ:sec:simulation}. The true $M_{\rm g}$ used in the
simulation is shown by the vertical line.}\label{SZ:fig:BL_UG_1d}
\end{center}
\end{figure}

For the Simulation `A' (with cluster and radio point sources), the
probability odds ratio for cluster detection $R$ was evaluated to be
$e^{12.2 \pm 0.2}$, showing an overwhelming evidence in favour of a `true'
cluster detection.  We plot the 2-D and 1-D marginalized posterior
distributions of the cluster parameters along with the true parameter
values used in the simulation in Figures \ref{SZ:fig:BL_tri} and
\ref{SZ:fig:BL_1d} respectively. In Figure \ref{SZ:fig:BL_1d}, we
also plot the prior distributions imposed on the parameters. The 
inferred cluster parameter means and 1-$\sigma$ uncertainties are 
listed in Table~\ref{tab:sz_correct_analysis}. It is clear from this 
table and the posterior plots that all the model parameters have been
estimated to reasonable accuracy. It can also be seen that the
posterior for the cluster temperature is largely unconstrained and the
expected degeneracy between $T$ and $M_{\rm g}$ is also
evident. Clearly some additional information on cluster $T$ is
required to get a sensible estimate for $M_{\rm g}$. This
information can come from the X-ray observation of the same cluster or
from an optical velocity dispersion measurement. $T$ calculated in
such a way can be used as a prior for the analysis of the cluster. The
result of a temperature measurement of $(8 \pm 2)$ keV applied as a
Gaussian prior on $T$ is shown in Figure \ref{SZ:fig:BL_UG_2d} and
\ref{SZ:fig:BL_UG_1d}.
\begin{table}
\begin{center}
\begin{tabular}{cc}
\hline
Parameters & Inferred values \\
\hline
$x_c$/arcsec & $12 \pm 11$ \\
$y_c$/arcsec & $3 \pm 10$ \\
$r_c/h^{-1}$kpc & $276 \pm 177$ \\
$\beta$ & $0.8 \pm 0.3$ \\
$M_{\rm g}/h^{-2}M_{\sun}$ & $(1.5 \pm 0.5) \times 10^{14}$ \\
$T$/keV & $10 \pm 5$ \\
\hline
\end{tabular}
\caption{Inferred cluster parameters values for the analysis with
cluster Simulation `A' discussed in Section \ref{SZ:sec:simulation}. All
noise sources discussed in Section~\ref{SZ:sec:noise} were included in 
the covariance matrix and the observed radio sources were treated as 
nuisance parameters. The probability odds ratio for cluster detection 
$R$ was evaluated to be $e^{12.2 \pm 0.2}$.}
\label{tab:sz_correct_analysis}
\end{center}
\end{table}

It is also of interest to investigate the importance of modelling
the three radio point sources, with properties given in Table
\ref{sz:tab:sim}, that were added to Simulation `A'. Assuming
known source positions and allowing the source fluxes to vary,
results in a massive increase in evidence over that for an analysis
(incorrectly) assuming no radio point sources.
\begin{equation}
\frac{P(\mathbfit{D}|3\,\mathrm{sources})}{P(\mathbfit{D}|\mathrm{No}\,\mathrm{sources})}=e^{961.9 \pm 0.2}.
\label{SZ:eq:sz_source_comp}
\end{equation}
We list the inferred cluster parameters for the analysis (incorrectly)
assuming no radio point sources in Table
\ref{tab:sz_noSrc_analysis}. Comparing the parameter values in Table
\ref{tab:sz_noSrc_analysis} with the values used in the simulation, it
is clear that the model used in the analysis is incorrect. The
algorithm is trying to fit for the cluster assuming no radio point
sources in the field, but the presence of radio point sources is
forcing the algorithm to compensate by preferring larger cluster $r_c$ and
$\beta$. This highlights the importance of radio point souce
information while analyzing the SZ data.
\begin{table}
\begin{center}
\begin{tabular}{cc}
\hline
Parameters & Inferred values \\
\hline
$x_c$/arcsec & $14 \pm 10$ \\
$y_c$/arcsec & $24 \pm 8$ \\
$r_c/h^{-1}$kpc & $546 \pm 65$ \\
$\beta$ & $1.4 \pm 0.1$ \\
$M_{\rm g}/h^{-2}M_{\sun}$ & $(2.1 \pm 0.8) \times 10^{14}$ \\
$T$/keV & $11 \pm 5$ \\
\hline
\end{tabular}
\caption{Inferred cluster parameters values for the analysis with
cluster Simulation `A' discussed in Section \ref{SZ:sec:simulation}
and (incorrectly) assuming no observed radio point sources.}
\label{tab:sz_noSrc_analysis}
\end{center}
\end{table}

Finally, to investigate the importance of including the noise
contribution due to unsubtracted radio sources, we analysed Simulation
`A' again but this time without including the contribution from the
confusion noise to the covariance matrix. The priors on the cluster
and radio source parameters used were the same as listed in
Table~\ref{tab:cluster_priors}. This resulted in the probability odds
ratio for cluster detection $R$ of $e^{50.8 \pm 0.2}$, again showing a
very strong evidence in favour of a true cluster detection. We list
the inferred cluster parameters for this analysis in Table
\ref{tab:sz_noConf_analysis}. It can be seen from this table that
while most of the cluster parameters have been inferred to reasonable
accuracy, the inferred cluster position is 5$\sigma$ away from the
true centre.
\begin{table}
\begin{center}
\begin{tabular}{cc}
\hline
Parameters & Inferred values \\
\hline
$x_c$/arcsec & $13 \pm 10$ \\
$y_c$/arcsec & $40 \pm 8$ \\
$r_c/h^{-1}$kpc & $296 \pm 175$ \\
$\beta$ & $0.8 \pm 0.3$ \\
$M_{\rm g}/h^{-2}M_{\sun}$ & $(1.6 \pm 0.5) \times 10^{14}$ \\
$T$/keV & $10 \pm 5$ \\
\hline
\end{tabular}
\caption{Inferred cluster parameters values for the analysis with
cluster simulation `A' discussed in Section \ref{SZ:sec:simulation}
and (incorrectly) assuming no unsubtracted radio sources below
$S_{\mathrm{lim}}$.}\label{tab:sz_noConf_analysis}
\end{center}
\end{table}
%

\section{Conclusions}\label{SZ:sec:conclusions}

Extracting and parametrizing clusters in SZ data is an extremely
challenging task due to the presence of (both observed and unsubtracted)
contaminating radio sources and primary CMB anisotropies. We have
described an efficient approach using the {\sc MultiNest} algorithm to
model galaxy clusters in multi-frequency pointed SZ observations, in
the presence of radio point sources. The parameters of the radio
sources are treated as nuisance parameters, which allows for the
fitting of these parameters simultaneously with the cluster parameters
and consequently for the uncertainties in the measurements of the
radio source parameters to be propagated to the cluster parameter
inferences. We considered the three main sources of noise for SZ
observations: (a) receiver noise, (b) primary CMB anisotropies
and (c) confusion noise due to unsubtracted radio sources. We have
also shown that it is extremely important to take into account all
these noise contributions in the analysis to get the correct posterior
distributions for the cluster parameters. Even with this extra
complexity, we are able to analyse a pointed SZ cluster observation
with six frequency channels on a single Intel Woodcrest 3.0-GHz
processor in about five hours. The work presented in this paper is
limited to pointed cluster observations. We plan to extend this
methodology to multi-field survey observations in a future study. The
code is fully parallel, making our algorithm a viable option for even
the deepest SZ surveys.

Our analysis methodology is easily extendable to do a joint analysis
of clusters using SZ, lensing and X-ray data-sets. This data fusion is
extremely important for understanding cluster physics, as there are
several degeneracies in cluster parameters when modelled through the
SZ effect only, the most significant of which are between $r_c$ and
$\beta$, and between $T$ and $M_{\rm g}$. Better constraints on
these parameters can be obtained by using information from different
waveband observations.

\section*{Acknowledgments}

This work was carried out largely on the Darwin Supercomputer of the
University of Cambridge High Performance Computing Service
(\texttt{www.hpc.cam.ac.uk}), provided by Dell Inc.~using Strategic
Research Infrastructure Funding from HEFCE. The authors would like to
thank Stuart Rankin for computing assistance. The work was conducted
in cooperation with SGI/Intel using the Altix 3700 supercomputer at
DAMTP, University of Cambridge supported by HEFCE and STFC, and we are
grateful to Andrey Kaliazin for his computing assistance. We thank
Phil Marshall, Katy Lancaster, Michael Bridges and members of the AMI
Consortium for useful discussions. FF is supported by the Cambridge
Commonwealth Trust, the Cambridge Isaac Newton Trust and the Pakistan
Higher Education Commission Fellowships.

\bibliographystyle{mn2e}
\bibliography{mcadam}

\bsp

\label{lastpage}


\end{document}